\documentclass[journal]{IEEEtran}
\IEEEoverridecommandlockouts
\usepackage{graphicx}
\usepackage{epstopdf}
\usepackage{amsmath}
\usepackage{amssymb}
\usepackage{amsthm}
\usepackage{epsfig}
\usepackage{times}
\usepackage{setspace}
\usepackage{bm}
\usepackage{siunitx}
\usepackage{threeparttable}
\usepackage{balance}
\usepackage{tabularx}
\usepackage{subcaption}
\usepackage{xcolor}
\usepackage{fancyhdr}
\usepackage{diagbox}
\usepackage{array}
\usepackage{float}

\usepackage{textcase}
\usepackage[tablename=TABLE]{caption}
\DeclareCaptionTextFormat{up}{\MakeTextUppercase{#1}}
\usepackage[T1]{fontenc}
\usepackage{times}
\usepackage[mathscr]{eucal}
\usepackage{wrapfig}
\usepackage{multirow}
\usepackage{tablefootnote}
\usepackage{makecell}

\newcommand{\equals}{\!=\!}
\newcommand{\lteq}{\!\le\!}
\newcommand{\minus}{\!-\!}
\newcommand{\gteq}{\!\ge\!}
\newcommand{\gthan}{\!>\!}
\newcommand{\lthan}{\!<\!}
\newcommand{\plus}{\!+\!}

\newcommand{\F}{\mathrm{F}}
\newcommand{\e}{\mathrm{e}}

\newcommand{\E}{\mathrm{E}}
\newcommand{\dx}{\:\mathrm{d}}

\renewcommand{\d}{\mathbf{d}}
\newcommand{\R}{\mathcal{R}}
\newcommand{\bE}{\overline{\mathcal{E}}}
\newcommand{\pr}{\mathrm{pr}}
\newcommand{\pl}{\mathrm{pl}}
\newcommand{\fr}{\mathrm{fr}}

\newcommand{\cF}{\mathcal{F}}
\newcommand{\TSP}{\mathcal{S}}

\newcommand{\cT}{\mathcal{T}}

\DeclareMathOperator*{\argmax}{argmax}
\DeclareMathOperator*{\argmin}{argmin}

\begin{document}
\title{Balancing the Energy Consumption and Latency of Over-the-Air Firmware Updates in LoRaWAN}
\author{ Siddhartha S. Borkotoky, \IEEEmembership{Member, IEEE} 
\thanks{
The author is with the Indian Institute of Technology Bhubaneswar, Khordha 752050, India. (e-mail: borkotoky@iitbbs.ac.in)}
}

\maketitle
\thispagestyle{fancy}
\lhead{{\color{gray} This work appears in the IEEE Transactions on Industrial Informatics (DOI: 10.1109/TII.2025.3563539). This is the author's version, and posted for personal use, not for redistribution.}}

\begin{abstract}
Over-the-air firmware updates are crucial for mitigating security threats and maintaining up-to-date device functionality in Long Range Wide Area Networks (\mbox{LoRaWANs}). 
LoRaWAN end devices are usually energy-constrained, and LoRaWAN transmissions are subject to duty-cycle restrictions. Consequently,  controlling the energy expenditure and update-delivery latency of FUOTA are key challenges. 
We propose a flexible scheme that achieves a tunable trade-off between the energy consumption and delivery delay. The scheme employs the LoRa spreading factors sequentially to transmit update-carrying frames, sending a fixed number of frames with a given spreading factor before moving to the next. By adjusting the smallest spreading factor to be used and the number of transmissions per spreading factor, a suitable energy-delay trade-off can be achieved. Thus, time-sensitive updates, such as security patches, may be sent with a low-delay-high-energy setting, whereas a more energy-efficient but higher-delay setting may be used for non-critical updates.
\end{abstract}


\vspace{-2mm}
\section{Introduction}
\label{sec:intro}
LoRaWANs support low-rate, long-distance, and low-power transmissions of small amounts of data by end devices (EDs), and exhibit lifetimes of up to several years. In a long-lived network, it is necessary to occasionally update the firmware at the EDs to enhance security features or implement new functionalities. 
Since a LoRaWAN can feature many EDs, potentially in hard-to-reach locations, it is beneficial to update them wirelessly without manual intervention. This paradigm, referred to as Firmware Update Over The Air (FUOTA), has been standardized by the LoRa Alliance's FUOTA working group~\cite{FUOTA}. 
In FUOTA, a firmware update server instructs the LoRaWAN gateway to transmit the firmware image to all eligible EDs. However, this process involves multiple challenges.  First, LoRaWAN transmitters must comply with strict duty-cycle regulations. For example, in the European Union, the maximum duty cycle must not exceed 1\%. Given the substantial size of a typical firmware image, duty-cycling has the effect that the gateway needs a long time to deliver the complete update to an ED. If the EDs are updated sequentially, it would take an unreasonably long time to update all devices. Therefore,  FUOTA specifications suggest simultaneous delivery to all recipients by exploiting the broadcast nature of the wireless medium. Secondly, owing to its size, the update must be split into many fragments, such that each fragment fits into a single link-layer frame. Keeping track of the fragment delivery status at each recipient and retransmitting undelivered fragments via an acknowledgment-based mechanism can be extremely complex in a massive network. Therefore, FUOTA specifications advocate the use of forward erasure correction (FEC). In FEC, redundant fragments are transmitted to enable an ED to compensate for lost fragments without explicitly requesting retransmissions~\cite{FUOTAFEC}. Yet another challenge, which is the focus of this work,  involves the selection of physical-layer parameters for multicast transmissions. The noise tolerance of LoRaWAN transmissions depends on a parameter called the spreading factor (SF), which takes integer values between 7 and 12~\cite{GeR17}. Higher SFs provide better receiver sensitivity, meaning that frames can be received over longer distances. However, higher SFs increase the frame duration, which in turn increases the time needed to deliver the update and the energy expenditure at the transmitter and receiver. With the EDs distributed over a large region, different SFs may be appropriate for different devices. Most existing works on FUOTA assume that the gateway uses an SF that is adequate even for the farthest device~\cite{AFM20, ACC20, NNA22, SNY23, SNY24}. The authors of~\cite{MZK23} show that this approach is not energy efficient for nearby devices. They propose an energy-efficient mechanism in which the EDs are grouped according to their optimal SFs. The update is delivered separately to each group. However, this approach incurs large delivery delays, since a group begins receiving frames only after all previous groups have been updated. 

To address the limitations of the fixed-SF and group-based approaches, we propose a transmission scheme that provides fine-grain control over the energy-latency trade-off. The scheme employs multiple SFs, with a fixed number of transmissions assigned to each SF. This number, along with the set of SFs, can be adjusted to control the energy-latency trade-off according to the needs of the update application. The scheme provides large reductions in the update time compared to the group-based scheme. Its energy consumption is only slightly higher than that of the group-based scheme and significantly lower than that of the fixed-SF scheme.             
\vspace{-3mm}
\subsection{Related Works}
In~\cite{AFM20}, the authors analyzed the impact of LoRaWAN transmission settings on FUOTA performance. It is shown that the higher data-rates (i.e., lower SFs) facilitate faster update and lower energy consumption at the cost of reduced update efficiency (i.e., the fraction of devices that receive the complete update). This observation underscores the need for careful SF selection. The trade-off between data rate and update efficiency is also highlighted in~\cite{ACC20}. A practical evaluation of FUOTA  is reported in~\cite{NNA22}. 
Efficient information delivery in wireless networks requires a mechanism to compensate for inevitable frame losses. For this, the FUOTA working group proposed an FEC scheme~\cite{FUOTAFEC} that transmits redundant coded fragments (linear combinations of the original fragments) using a low-density parity-check (LDPC) code. The redundant fragments may enable an ED to correct for some lost data. A related FEC technique called rateless coding is used in~\cite{SNY23}. A salient feature of a rateless code is that it allows the source to continue sending redundancy until all recipients recover all lost data. In contrast, the LDPC mechanism of~\cite{FUOTAFEC} transmits only a predefined amount of redundancy, hence some recipients may fail to recover all lost data. In~\cite{SNY23}, the gateway multicasts rateless coded fragments up to a predefined time. The nodes that have yet to receive the complete update are then individually delivered the remainder of the update using a beamforming strategy. The transmission scheme of~\cite{SNY23} is improved in~\cite{SNY24} by incorporating a security mechanism.

The aforementioned techniques employ a fixed SF for all transmissions. However, fixed-SF assignments can be inefficient. If the SF is large, nodes near the gateway spend more energy by receiving long frames  although they could successfully receive shorter frames. With a small SF, nearby nodes spend less energy, but distant nodes experience excessive frame losses and require very large update times. As a remedy,~\cite{MZK23} suggests grouping the nodes according to their appropriate SFs. The update is delivered group-wise. Once all nodes using a certain SF receive the update, transmissions to the next group begin. 
However, as mentioned earlier, the sequential handling of groups increases the overall delivery time.

\vspace{-3mm}
\subsection{Contributions of this Paper}
In contrast to fixed-SF and group-based schemes, which aim for either low energy consumption or low delay, we present a flexible transmission scheme that can tune the trade-off between the two metrics. By adjusting a pair of design parameters, the operating point can be set as per the need of the update application (e.g., a high-priority update can employ less energy-efficient operation for faster completion, and vice versa). The key features of the proposed scheme are as follows:
\begin{itemize}
    \item The gateway employs the SFs sequentially -- sending a fixed number of frames with a given SF, then the same number of frames with the next higher SF, and so on.  
    \item The EDs try to receive all frames, regardless of the SF. Hence, unlike in the fixed-SF scheme, nearby EDs are not forced to receive exclusively over an unnecessarily large SF. This saves energy at such EDs. Simultaneously, the use of larger SFs ensure delivery to the distant EDs. Unlike in the group-based scheme, moving to the next higher SF does not depend on the update status of any ED. This reduces the overall update time. 
    \item The SF set and the number of transmissions per SF can be adjusted to achieve the desired trade-off between the energy expenditure and the update time. We present an analytical framework to characterize the energy and delay performance as a function of these parameters, so that appropriate parameter values can be chosen depending on application requirements.   
\end{itemize}

The remainder of the paper is organized as follows. Section~\ref{sec:background} provides a brief background on LoRaWAN, FUOTA, and FEC. The proposed scheme is described in Section~\ref{sec:tx_policy}. Section~\ref{sec:system_model} describes the system model used for analytical modeling. The analytical framework is described in Section~\ref{sec:perf_analysis}. The benchmark schemes for performance evaluations are described in Section~\ref{sec:benchmarks}. In Section~\ref{sec:res}, numerical results are presented. Finally, Section~\ref{sec:conlusion} concludes the paper.

\section{Background}
\label{sec:background}

\subsection{LoRaWAN}

LoRaWANs support star-topology sensor networks, with  single-hop communication between EDs and the gateway. The EDs transmit frames on the uplink using pure ALOHA. A frame comprises a preamble for synchronization, a payload portion encoded with a channel code, and an optional header. The frame duration   depends on the SF, which takes integer values between 7 and 12. The airtime of a frame having SF $i$ and carrying $b$ bytes of payload is 
\begin{align}
    l_i^{(\fr)}(b) = l_i^{(\pr)} + l_i^{(\pl)}(b),
\end{align}
where $l_i^{(\pr)}$ and $l_i^{(\pl)}(b)$ are the duration of the preamble and the payload portions, respectively. For a transmission bandwidth $\mathrm{BW}$, the preamble duration is given by
\begin{align}
    l_i^{(\pr)} = (n_{\mathrm{pr}} + 4.25) \frac{2^{i}}{\mathrm{BW}}.
\end{align}
The payload duration is
\begin{align} 
    l_i^{(\pl)}(b) \equals \bigg[8 \plus \max\left\{ \left\lceil \frac{2b \minus i \minus 5h \plus 11}{i \minus 2y} \right\rceil (c \plus 4), 0\right\}\bigg]\,\frac{2^{i}}{\mathrm{BW}}, 
\end{align}
where  $n_{\mathrm{pr}}$ is the number of preamble symbols, $h$ is 0 or 1 depending on whether an optional header is included or not, $y$ is 1 when low data rate optimization is enabled and 0 otherwise, and $c$ takes integer values between 1 and 4 depending on the channel code~\cite{Sem13}. 

The receiver sensitivity improves with increasing SF.  Let  $\zeta_i$ be  the sensitivity for SF $i$. Then $\zeta_7 \gthan \zeta_8 \gthan \cdots \gthan \zeta_{12}$. A frame with SF $i$ and received power $R$ is lost if \mbox{$R \lthan \zeta_i$}. In addition, a frame may be lost due to collision with interfering frames.  For analyzing collisions, we employ the often-used dominant-interferer model~\cite{GeR17}. In this model, a frame with SF $i$ and received power $R$ is lost upon collision with a frame using SF $i'$ and received power $R'$ if and only if $R/R' \lthan \xi_{i,i'}$, where $\xi_{i,i'}$ is the capture threshold. 

For  receptions on the downlink, three classes of operation are defined. In \mbox{class A}, an ED opens two receive windows following an uplink transmission to scan for potential downlink frames; at all other times, the ED is in the idle state.  In Class B, the ED opens additional receive windows (called ping slots) at predefined instants. The gateway periodically transmits beacon signals to synchronize the ED's ping slot boundaries with its downlink transmission instants. In Class C, the ED is always in the receive mode except when transmitting.

\subsection{FUOTA for LoRaWAN}
The LoRa Alliance's FUOTA working group has so far released five sets of specifications~\cite{FUOTAFAQ}.
Here we briefly mention the key features relevant to this work. For an in-depth discussion, the reader is referred to~\cite{AFM20}. The FUOTA process is handled by a \textit{firmware update server} with the help of a \textit{firmware update management entity}. The firmware image to be sent is divided into \textit{fragments} such that each fragment fits into a single LoRa frame. During the setup phase, a multicast group is formed comprising the EDs that need to be upgraded to the same firmware version. The fragments are than transmitted via wireless broadcast to these EDs by a gateway. The specifications include a mechanism for synchronizing the receptions at the EDs with the gateway's transmissions. Despite Class A being the most common mode of operation in LoRaWANs, FUOTA specifications require that the EDs be in Class B or C during the multicast session. This is because downlink transmissions in Class A can be only initiated following uplink transmissions, which makes it unsuitable for downlink-heavy FUOTA. Hence, Class A devices switch to Class B or C (as instructed via a control message during the set up phase) at the start of the FUOTA session, and return to Class A once the session ends.

\subsection{Forward Erasure Correction: Fixed-Rate vs. Rateless}
FUOTA specifications suggest transmitting coded redundancy to compensate for data loss that occurs due to noise and interference~\cite{FUOTAFEC}.  Suppose the firmware image comprises $k$ fragments (hereafter referred to as \textit{source fragments}). These  are mapped to $k'$
\textit{coded fragments} ($k'\gthan k$). Each coded fragment is the bit-wise XOR sum of certain source fragments chosen according to a LDPC encoding matrix~\cite{FUOTAFEC}. The  coded fragments are transmitted to the EDs rather than the  source fragments, resulting in $k \minus k'$ redundant transmissions. Upon  receiving any set of coded fragments containing $k$ linearly independent combinations of source fragments, the ED runs a decoding algorithm to determine the source fragments. This eliminates the need for EDs to send individual acknowledgements for the received fragments. Instead, an ED continues receiving frames until it collects \textit{any} set of coded fragments that enables successful decoding. 

An alternative to the LDPC code of~\cite{FUOTAFEC} is a rateless code. Here the value of $k'$ is not fixed. Instead, the gateway continues transmitting coded fragments until all EDs recover the image. Each coded fragment is produced by XOR-ing randomly selected source fragments. Well-known rateless codes include LT codes~\cite{Lub02} and raptor codes~\cite{Sho06}. A rateless code suitable for FUOTA over LoRaWAN is proposed in~\cite{SNY23}. The benefit of rateless coding over the fixed-rate LDPC code of~\cite{FUOTAFEC} is that in the former, the EDs that do not receive $k$ linearly independent coded fragments out of the $k'$ transmissions cannot recover the complete firmware image. In contrast, rateless codes can generate as many coded fragments as needed to guarantee update-completion success at all recipients. For this reason, we assume the use of a rateless code in this paper. However, the proposed method can be adapted for use with other types of codes, including LDPC codes.

\section{Proposed Transmission Scheme}
\label{sec:tx_policy}
We assume that the multicast group of intended recipients of the firmware update has been formed, and the necessary synchronization has been performed as per FUOTA specifications. The basic operation of our method is as follows: The gateway transmits a stream of rateless coded FUOTA fragments until all recipients recover the complete firmware image. The initial transmissions use a  small SF. These transmissions permit nearby recipients to recover the image with low energy expenditure. To serve the more distant EDs,    
the SF is then gradually increased. The method is described in detail below.

Let $b$ denote the number of bytes in a \textit{data block}, which represents either the firmware image or its compressed version~\cite{SNY23}. The data block is divided into $k$ equal-sized \textit{source fragments} $\d_0, \d_1,\ldots,\d_{k-1}$. If $b$ is not an integer multiple of $k$, then the data block is zero padded to make it so. Therefore, the number of bytes per information block is
\begin{align}
    b_k = \left\lceil b/k \right\rceil,
\end{align}
where $\left\lceil x \right\rceil$ is the smallest integer equal to or greater than $x$. The sender then applies a rateless code to the source fragments to produce a sequence of \textit{coded fragments}.

Each coded fragment is encapsulated in a LoRaWAN frame and broadcast to the recipients. The gateway employs the set of SFs $\{L,L \plus 1,\ldots,M\}$ (\mbox{$7 \leq L \leq M \lteq 12$}) in a sequential manner as follows: Let the positive integer $w$ be a design parameter. The gateway transmits the first $w$ frames with SF $L$, the second  $w$ frames with SF $L \plus 1$, and so on, until SF $M$ is reached. All subsequent transmissions until update delivery completion at all recipients are made with SF $M$.

Following standard reception procedure, a recipient attempts to acquire the preamble of a downlink frame. If acquisition fails, the recipient does not attempt to demodulate the remainder of the frame. After successful preamble acquisition, demodulation is attempted. The successfully received coded fragments are then stored in memory.  Upon reception of the complete update (i.e., after  retrieving all source fragments by successfully decoding the rateless code), the recipient transmits an acknowledgement to the gateway.

\vspace{-3mm}
 \section{System Model}
\label{sec:system_model}  

For our performance analysis, the firmware recipients are assumed to be uniformly distributed in a circular region of radius $\R_0$, with the gateway at the center. 
The recipients are interspersed between other transmitting devices referred to as \textit{interferers}, possibly belonging to many different networks and distributed uniformly over an infinitely large area according to a Poisson point process (PPP) of intensity $\lambda_I$. Each interferer generates messages according to a Poisson arrival process with a mean of $\lambda_f$ frames per second, and makes transmissions using pure ALOHA. For simplicity, all transmissions (from the interferers as well as the gateway performing FUOTA) are assumed to be made with power $p_t$. The interferer payload sizes and consequently frame lengths are random. We denote by $\overline{l}_j$  the average duration of an interfering frame having SF $j$. A total of $n_f$ non-overlapping frequency bands are available to all transmitters. The frequency band to be used by the gateway is known a priori to the intended recipients, whereas an interfering frame is equally likely to use any one of the $n_f$ bands. Likewise, the SF to be used by the gateway for a FUOTA frame is known to the recipients, whereas the SF of an interfering frame is random. We use $\eta_j$ to denote the probability that an interferer uses SF $j$, where $7 \lteq j \lteq 12$.

\begin{figure}
    \centering
    \includegraphics[scale=0.25]{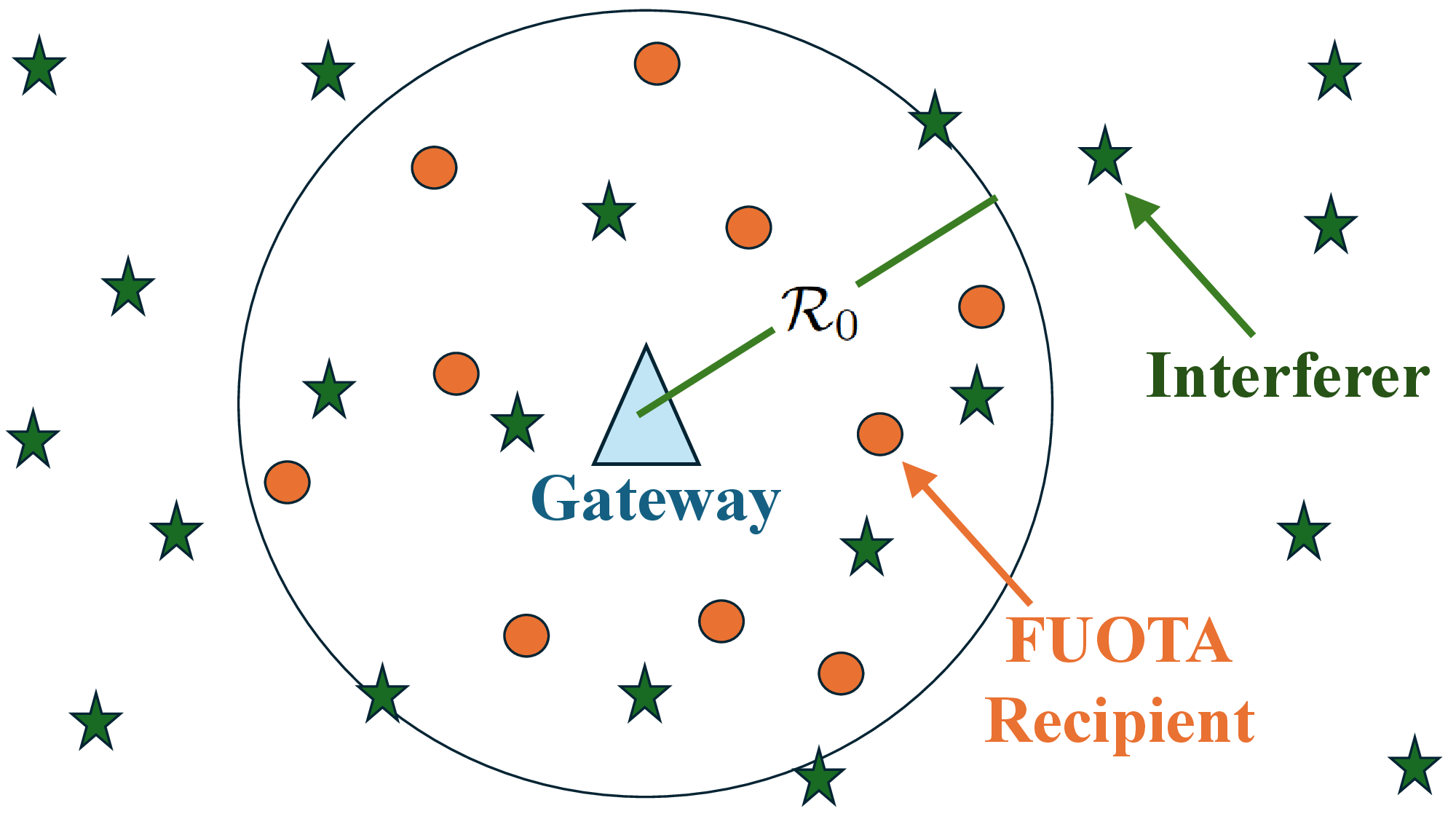}\caption{Model of the network.}\
    \vspace{-4mm}
    % setlength\belowcaptionskip{-10mm}
    \label{fig:network_model}
\end{figure}

The links experience path loss with exponent $\alpha$ and independently and identically distributed (i.i.d.) block Rayleigh fading. For transmit power $p_t$, the received power at distance $d$ is  \mbox{$R \equals\gamma_0 p_t A d^{-\alpha}$}, where $A$ is the fading coefficient with pdf $f_A(a) \equals \e^{-a}$ for $a \gteq 0$.  The quantity $\gamma_0$ is given by $\gamma_0 \equals g_tg_r  L / 4\pi$, where $g_t$ and $g_r$ are transmit and receive antenna gains, respectively, and $L$ is the signal wavelength.

\section{Analytical Framework}
\label{sec:perf_analysis}
We wish to find the energy spent by an ED to receive the update and the total time required to receive it.  Let $\bE_T(d_0)$ be the total energy expenditure at an ED at distance $d_0$ from the gateway, $\bE(d_0)$ be the energy spent in receiving FUOTA fragments, and $\mathcal{E}_C(d_0)$ be the energy spent on control  overheads such as receiving  synchronization beacons and transmitting acknowledgments. Let $l_c$ be the total duration of downlink control frames received, and $l_a$ be the duration of an acknowledgment frame. Let $p_r$ and $p_t$ be the power consumption during demodulation and transmission, respectively. Let $\overline{\cT}(d_0)$ be the average time the recipient needs to receive the complete update, $\TSP^{(\fr)}_i(d_0)$ be the probability that a recipient at distance $d_0$ from the gateway is able to successfully receive a frame with SF $i$, $\cF_i^{(\pl)}$ be the probability that a frame's payload is received in error conditioned on successful acquisition of the preamble, and $\cF_i^{(\pr)}$ be the probability that preamble acquisition fails.  Define $\mathcal{S}_{i}^{(\pl/\pr/\fr)} \equals 1 \minus \cF_{i}^{(\pl/\pr/\fr)}$. 
Let $e_i^{(\pr)}$ and $e_i^{(\fr)}$ denote the energy spent in demodulating the preamble and the complete frame, respectively. Hence, $e_i^{(\fr)} \equals p_r l_i^{(\fr)}(b_k)$ and    $e_i^{(\pr)} \equals p_r l_i^{(\pr)}$. Let $\overline{N}_s$ denote the average number of coded fragments required to  recover the firmware image (derived in Appendix~\ref{app:Ns}.)
 
While our system model assumes the interferers to be distributed over an infinite region, transmissions from very distant nodes are unlikely to cause interference at a recipient. For analytical convenience, we define the \textit{interference radius} $\R_I$ to be the maximum distance between a recipient and an interferer such that the received power at the recipient from the interferer exceeds the recipient's maximum receiver sensitivity (i.e., $\zeta_{12}$) with a small predefined probability $\delta$. Since received power at distance $d$ is given by $\gamma_0 p_t A d^{-\alpha}$,
\begin{align} \nonumber
\label{eq:RI}
    \R_I &= \argmax_{d} \{P(\gamma_0 p_t A d^{-\alpha} > \zeta_{12}) \gteq \delta \} \\
    &= \argmax_{d} \{\exp(-(\gamma_0 p_t)^{-1} d^{\alpha}\zeta_{12}) \gteq \delta \}.
\end{align}
The practical interpretation of~\eqref{eq:RI} is as follows: Consider a colliding frame from an interferer located beyond a distance $\R_I$ from a given recipient receiving in its most sensitive setting (SF 12). The probability that this frame has enough power to be detected by the recipient is no greater than $\delta$. For lower SFs, the probability is even lower. 
Since enforcing a frame loss requires that the interference power be at least $\xi_{j,j'}$ dB above the desired signal, the probability of devices located beyond distance $\R_I$ causing frame losses at the recipient is negligible. Hence, we ignore such devices. Instead, intereferers are approximated as being spread over a circle of radius $\R_I$, with the intended recipient at the center.  A smaller $\delta$ implies a lower approximation error. We employ $\delta \equals 0.01$, since we have found this choice to result in no noticeable loss of accuracy. 

Since the interferers are uniformly distributed, the pdf for the distance between the ED  and an arbitrary interferer within the interference radius is 
$
    f_D(u) \equals
        2u/{\R^2_I}, \quad u \lthan \R_I.
$
The expected number of interferers within the interference radius is $\lambda_I \pi \R_I^2$. Hence, for any recipient, the probability that there are $n$ interferers within the interference radius is
\begin{equation} 
    P_I(n) = (\lambda_I \pi \R_I^2)^{n} \e^{-\lambda_I \pi \R_I^2}/n!.
\end{equation}

\vspace{-3mm}
\subsection{Average Energy Expenditure}
As per notation introduced at the beginning of this section, $\bE_T(d_0)\equals \bE(d_0) \plus \mathcal{E}_C(d_0)$, where $\mathcal{E}_C(d_0)\equals p_r l_c \plus p_t l_a$, Hence, our primary task is to derive $\bE(d_0)$. Its value depends on the amount of interference in the network. Higher interference implies more frame losses, and consequently more energy spent in unsuccessful reception attempts.  
Since the intensity of interference depends on the number of interferers, we  write
\begin{align}
    \bE(d_0) = \sum_{n=0}^{\infty} P_I(n) \bE(d_0|n),
\end{align}
where $\bE(d_0|n)$ is the expected energy expenditure conditioned on there being $n$ interferers within the interference radius. 

To derive $\bE(d_0|n)$, first recall that  transmissions occur in rounds, with the first $w$ transmissions using SF $L$, the next $w$ using SF $L \plus 1$, and so on. For simplicity, let us refer to the first round as round $L$, the second round as round $L \plus 1$, and so on. Once round $M$ is over, all subsequent  transmissions (which continue using SF $M$)  are interpreted as belonging to round $M \plus 1$. Let $m_0$ denote the round during which the recipient receives the coded fragment that results in decoding success.   Let ${\bE}^{(\fr)}_i(d_0)$ be the recipient's average energy consumption per frame-reception attempt in round $i$. When $i \lthan m_0$, the recipient makes exactly $w$ frame reception attempts, spending a total energy of $w{\bE}^{(\fr)}_i(d_0)$. Let $\eta(d_0)$ be the expected number of reception attempts made by the recipient in the $m_0$-th round. Thus, the average energy spent in this round is $\eta(d_0){\bE}^{(\fr)}_{m_0}(d_0)$. Summing over rounds $L$ through $m_0$, the total energy expenditure is 
\begin{align}
\label{eq:Edn}
    \bE(d_0|n) = w\sum_{i=L}^{m_0-1}{\bE}^{(\fr)}_i(d_0) +  \eta(d_0){\bE}^{(\fr)}_{m_0}(d_0).
\end{align}
Note that the terms $m_0$, $\bE^{(\fr)}_i$, $\bE^{(\fr)}_{m_0}$, and $\eta$  on the RHS depend on $n$. For brevity, we have dropped the conditioning notation.

To determine $m_0$, first consider the two possibilities: The recipient may complete the decoding within round $M$ (each of which has $w$ transmissions); else, round $M \plus 1$ must be initiated. The first scenario occurs if $\sum_{i=L}^{M}\overline{R}_i(d_0) \gteq \overline{N}_s$, where $\overline{R}_i(d_0)$ is the expected number of successful receptions out of the $w$ transmissions in round $i$. Note that
\begin{align}
\overline{R}_i(d_0) = w\TSP^{(\fr)}_i(d_0).    
\end{align}
If $\sum_{i=L}^{M}\overline{R}_i(d_0) \lthan \overline{N}_s$, the second scenario (i.e., $m_0 \equals M \plus 1)$ occurs. Hence
\begin{align}
    m_0 = \begin{cases}
        \displaystyle{\argmin_{m}}\left\{\sum_{i=L}^{m}\overline{R}_i(d_0) \gteq \overline{N}_s \right\}, \quad &\displaystyle\sum_{i=L}^{M}\overline{R}_i(d_0) \gteq \overline{N}_s \\
        M+1, \quad &\text{otherwise}.
    \end{cases}
\end{align}

Since the recipient collects $\sum_{i=L}^{m_0-1}\overline{R}_i(d_0)$ fragments in the first $m_0 \minus 1$ rounds, it must receive  another 
$\overline{N}_s - \sum_{i=L}^{m_0-1}\overline{R}_i(d_0)$ fragments in round $m_0$. Therefore, the expected number of reception attempts made by the recipient round $m_0$ is 
\begin{align}
    \eta(d_0) = \frac{\overline{N}_s - \sum_{i=L}^{m_0-1}\overline{R}_i(d_0)}{1 - \TSP^{(\fr)}_{m_0}(d_0)}.
\end{align}

To determine ${\bE}^{(\fr)}_i(d_0)$,  note that a frame reception attempt has three possible outcomes: (1) The frame is correctly demodulated. (2) The preamble is acquired but the payload is incorrectly demodulated. (3) Preamble acquisition fails. In the first two instances, the recipient demodulates the entire frame, and hence incurs an energy expenditure of $e_i^{(\fr)}$.  In the third case, payload demodulation does not occur, and the energy expenditure is $e_i^{(\pr)}$. Accounting for the three possibilities, the average energy expenditure per frame can be expressed as 
\begin{align}
    {\bE}^{(\fr)}_i(d_0) = \TSP^{(\fr)}_i e_i^{(\fr)} + \cF^{(\pl)}_ie_i^{(\fr)} + \cF^{(\pr)}_ie_i^{(\pr)}. 
\end{align}
As derived in Appendix~\ref{app:S},
\begin{align} \nonumber
\label{eq:SI_full_pr}
    \cF_{i}^{(\pr)}(d_0) = &1-\int_{\zeta_i d_0^{\alpha}/\gamma_0 p_t}^{\infty}\Big(1 - \frac{2 \lambda_f}{\alpha\R_I^2 n_f }\sum_{j=7}^{M}\frac{\eta_j(l^{(\pr)}_i+\overline{l}_j)}{\beta_j^{2/\alpha}}\\
    & \gamma(2/\alpha,\beta_{j} \R_I^\alpha)  \Big)^{n}\e^{-a}\dx{a},
\end{align} 
\begin{align} \nonumber
\label{eq:SI_full_fr}
    \TSP_{i}^{(\fr)}(d_0) = &\int_{\zeta_i d_0^{\alpha}/\gamma_0 p_t}^{\infty}\Big(1 - \frac{2 \lambda_f}{\alpha\R_I^2 n_f }\sum_{j=7}^{M}\frac{\eta_j\left(l^{(\fr)}_i(b_k)+\overline{l}_j\right)}{\beta_j^{2/\alpha}}\\
    & \gamma(2/\alpha,\beta_{j} \R_I^\alpha)  \Big)^{n}\e^{-a}\dx{a},
\end{align}
and 
\begin{align}
    \cF_{i}^{(\pl)}(d_0) = 1 - \frac{\TSP_{i}^{(\fr)}(d_0)}{1-\cF_{i}^{(\pr)}(d_0)},
\end{align}
where \mbox{$\beta_j \equals \xi_{i,j}^{-1}ad_0^{-\alpha}$} and $\gamma(i,x) \equals \int_{0}^{x}t^{i-1}\mathrm{e}^{-t}\dx{t}$ is the lower incomplete Gamma function.

\subsection{Average Update Time}
Assuming that the gateway transmits at the maximum permitted  percentage duty cycle of $\mathrm{DC}_{\max}$, the time needed to make $\nu$ transmissions in round $j$ is  
\begin{align}
    \cT_i(\nu) = \frac{100}{\mathrm{DC}_{\max}} \times \nu l_i(b_k). 
\end{align}

The expected update time $\overline{\cT}(d_0)$ for the recipient is the time needed to complete the first $m_0 \minus 1$ rounds, each involving $w$ transmissions, plus the time needed for $\eta(d_0)$ transmissions in the $m_0$-th round. Conditioned on there being $n$ interferers within the recipient's interference radius,
\begin{align}
    \overline{\cT}(d_0|n) = \sum_{i=L}^{m_0-1}\cT_i(w) +  \cT_{m_0}(\eta(d_0)).
\end{align}
As before, our notation does not explicitly indicate the dependence of $m_0$ and $\eta(d_0)$ on $n$ for the sake of conciseness. Upon deconditioning, we obtain
\begin{align}
        \overline{\cT}(d_0) = \sum_{i=0}^{\infty} P_I(n) \overline{\cT}(d_0|n).
\end{align}

\vspace{-3mm}
\section{Benchmark Schemes}
\label{sec:benchmarks}
To evaluate the efficacy of the proposed method, we compare it with the benchmark schemes described below.

\paragraph{Fixed-SF} Similar to~\cite{SNY23,SNY24}, this scheme employs rateless-coded multicast with the same SF for all transmissions. We assume an ideal rateless code that allows the recipient to recover the complete update upon receiving any $k$ coded fragments. This provides a performance upper bound for the practical rateless code of~\cite{SNY23,SNY24}. In~\cite{SNY23,SNY24}, a beamforming procedure is used to complete the update delivery via unicast transmissions after partial delivery via multicast. Here, we  do not assume beamforming capabilities. Consequently, all transmissions are in the multicast mode.

\paragraph{Group-Based} This scheme is inspired by~\cite{MZK23}. The recipients are divided into groups, with each group assigned a unique SF. A given recipient receives only those frames that use its assigned SF. The gateway starts the session by multicasting to the group having the smallest SF. Only when all recipients in a group receive the complete update, transmissions intended to the next group start. We again assume an ideal code wherein a recipient can recover the complete update upon receiving any $k$ coded fragments. This is an idealization of the mechanism of~\cite{MZK23}, which is essentially a form of systematic erasure correction.  We consider two grouping strategies as described below. 

\textit{Energy-based}: A recipient is assigned the SF that leads to the lowest expected receiver energy expenditure for the update. Thus, the SF for a device at distance $d_0$ from the gateway is
\begin{align}
    i^*_{\mathrm{(en)}} = \argmin_{i \in \{7,8,\ldots,12\}} \left\{ \frac{k}{\TSP_i^{(\fr)}(d_0)} \times e_i^{(\fr)}(d_0) \right\}.
\end{align}

\textit{Latency-based}: A recipient is assigned the SF that provides the smallest expected  update time. Thus, the SF for a device at distance $d_0$ from the gateway is
\begin{align}
    i^*_{\mathrm{(lt)}} = \argmin_{i \in \{7,8,\ldots,12\}} \left\{ \frac{k}{\TSP_i^{(\fr)}(d_0)} \times \frac{100 l_i^{(\fr)}(b_k)}{\mathrm{DC}_{\max}}\right\}.
\end{align}

\section{Numerical Results}
\label{sec:res}
We simulate the transmission of a 10 kB firmware image to 100 recipients uniformly distributed over a 1 km radius around a gateway operating at 1\,\% duty cycle. The image is split into 200 fragments of 50 bytes each. Unless stated otherwise, the interferers are distributed with an intensity of $5\times 10^{-5}$ devices per m$^2$, each interferer transmits one frame every 10 minutes on average, transmissions have 14 dBm power, the path-loss exponent is 2.5, interfering frames carry 5-byte payloads,  eight 125-kHz channels are available, and each channel and each SF is equally likely to be used by an interferer. FUOTA specifications do not explicitly specify a value for the total airtime of downlink control signals. For illustration, we assume this duration to be $l_c \equals 60$ seconds. For the proposed scheme, the default parameter values are $L \equals 7$, $M \equals 12$, and $w \equals 300$.  The receiver sensitivities and capture thresholds are as given in~\cite{GeR17} and~\cite{MSG19}, respectively.  The simulations are performed in MATLAB. Each data point is the average of 100 simulation runs. We use FSF-$i$, GB-E, and GB-L to refer to the benchmark fixed-SF scheme with SF $i$, group-based scheme with energy-based grouping, and group-based scheme with latency-based grouping, respectively.

\begin{figure}
\begin{subfigure}{0.5\textwidth}
  \centering
  % include first image
  \includegraphics[scale=0.55]{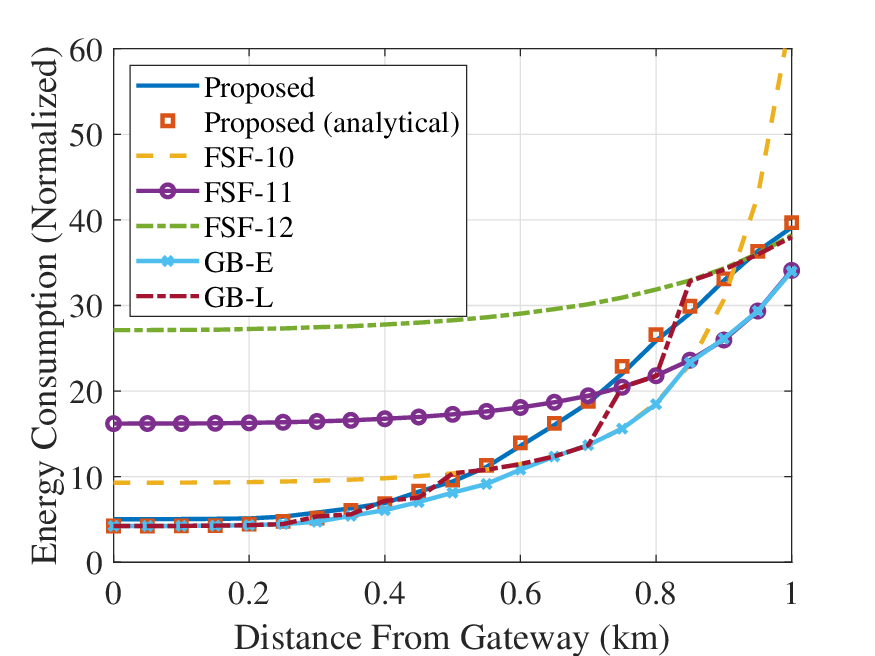}  
  \vspace{-1mm}
  \caption{Energy expenditure (Normalized).}
  \label{fig:energy_vs_dist}
\end{subfigure}
\begin{subfigure}{0.5\textwidth}
  \centering
  % include second image
  \includegraphics[scale=0.55]{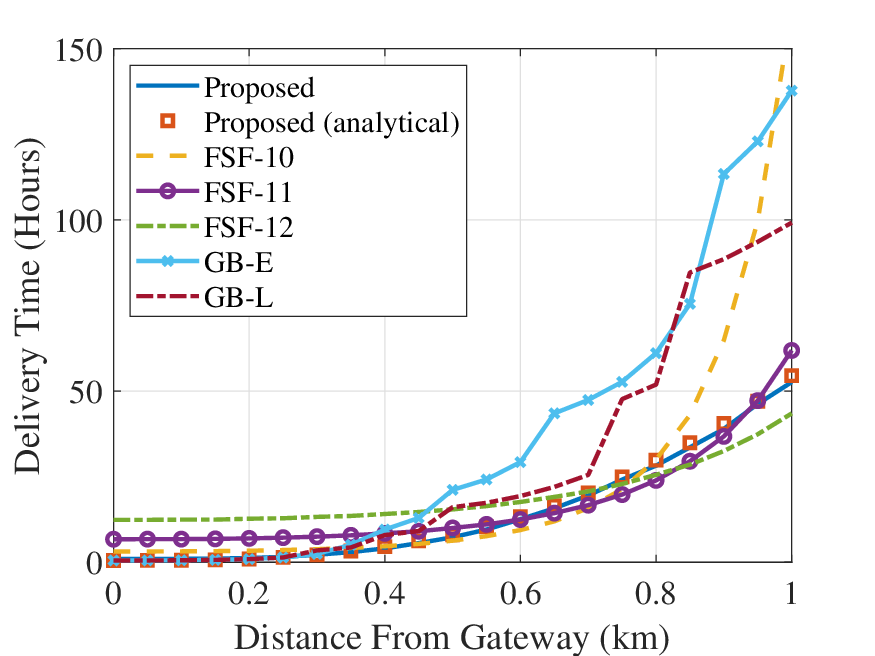}  
  \vspace{-1mm}
  \caption{Delivery time.}
  \label{fig:latency_vs_dist}
\end{subfigure}
% \vspace{-4mm}
% \setlength\belowcaptionskip{-6mm}
\caption{Performance as a function of node distance.}
\label{fig:vsDistance}
\end{figure}

The energy and delay performance of the different schemes as a function of the recipient's distance from the gateway are shown in Figs.~\ref{fig:energy_vs_dist} and~\ref{fig:latency_vs_dist}, respectively. The energy values are normalized via division by $kp_rl_f(7)$, which is the energy needed to receive the file using the shortest SF (7) over an erasure-free link. For FSF, results for SFs 10, 11, and 12 are shown, since SFs below 10 cannot reach the farthest recipient. We observe that while FSF outperforms the proposed scheme for certain distances, it is significantly inferior at other distances. For example, FSF-11 is more energy efficient than the proposed scheme for distant devices (with up to about 13\,\% lower energy expenditure), but spends about 200\,\% more energy at nearby devices. Likewise, FSF-12 provides about up to about 20\,\% lower delivery time at distant devices compared to the proposed scheme, but incurs up to 23 times larger delivery delays at nearby EDs, in addition to spending much higher energy at these EDs. For the group-based schemes, we observe that both provide slightly lower energy expenditure than the proposed scheme (up to about 20\,\% lower), but need much higher delivery times. Fig.~\ref{fig:vsDistance} also includes analytical results for the proposed scheme, demonstrating the accuracy of the derivation presented in Section~\ref{sec:perf_analysis}.

The impact of the aforementioned schemes on an ED's battery lifetime is of practical interest. Indeed, the exact lifetime depends on battery characteristics and the ED's transmission behavior. For a rough estimate, consider an ED that has battery capacity $C_b \equals 1200$ mAh, receives $N_u \equals 1$   firmware update per month  on average, and transmits a $b \equals 50$~byte uplink message in the unacknowledged mode every $T_m \equals 0.5$~hr using SF $s$. Using data from~\cite{CMV17}, we consider  current consumptions of $I_t \equals 83$~mA, $I_r \equals 38$~mA, and $I_s \equals 45$~$\mu$A during transmission, reception, and sleep states, respectively. Under ideal circumstances (linear battery behavior over time without state-of-health deterioration, ED sleeps at all times except when receiving FUOTA frames and transmitting measurements),  
the battery lifetime can be approximated by
\begin{align}
    L \approx \frac{C_b}{I_t\Delta_t + I_r\Delta_r + I_s\Delta_s} \quad \text{years},
\end{align}
where $\Delta_t$,  $\Delta_r$, and  $\Delta_s$ are the total amount of time (in hours) the ED transmits, receives, and sleeps, respectively, per year. It is easy to see that $\Delta_t \equals 365 \cdot (24/T_m) \cdot (\hat{l}_f/3600)$, where $\hat{l}_f$ is the uplink frame duration in seconds, $\Delta_r \equals 12N_uR_u$, where $R_u$ is the number of hours per year that the ED spends in receiving FUOTA frames (which can be found via simulations or inferred from the preceding analysis), and   $\Delta_s \equals 365\times 24 \minus (\Delta_t\plus \Delta_r)$. The expected battery lifetime and update delivery times for the proposed scheme for an ED at the network's edge ($d_0 \equals \mathcal{R}$, uplink SF 12) and an ED near the gateway ($d_0 \equals \mathcal{R}/4$, uplink SF 7)  are shown in Table~\ref{tab:lifetime}. For comparison, we choose the benchmarks FSF-11 and GB-E, because these two provide the best energy efficiency under the worst-case scenario of an ED being located at the network's edge. Table~\ref{tab:lifetime} shows that the proposed scheme provides comparable battery lifetimes vis-\`{a}-vis the benchmarks, and minimum or near-minimum delivery times for both node locations. By contrast, neither benchmark performs well in terms of both metric at both locations.

\begin{table}[t]
    \centering
       \caption{
Battery lifetime (LT) and update delivery time (DT)}
    \begin{tabular}{c|c|c|c|c|c|c}
      \multirow{2}{1em}{} & \multicolumn{3}{c|}{\small{$d_0\equals\mathcal{R}$, $s \equals 12$}}  & \multicolumn{3}{c}{\small{$d_0\equals\mathcal{R}/4$, $s \equals 7$}} \\ 
      & \footnotesize{Prop.} & \footnotesize{FSF-11} & \footnotesize{GB-E} & \footnotesize{Prop.} & \footnotesize{FSF-11} & \footnotesize{GB-E} \\\hline
    \small{LT (years)} & 1.42 & 1.47 & 1.47 & 1.82  &  1.66 & 1.82\\
    \small{DT (hours)} & 54.5 & 61.9 & 137.7 & 1.44 & 7.18 & 1.41\\ \vspace{-3mm} 
    \end{tabular}
    \label{tab:lifetime}

\end{table}

Table~\ref{tab:avgPerf} shows the energy expenditure and delivery delay averaged over all simulated distances. While GB-E and GB-L spend less energy than the proposed scheme, both incur much higher delivery times. The proposed scheme  outperforms all FSF schemes, both in terms of average energy and latency. The impact of varying interference levels is examined in Table~\ref{tab:traffic_impact} for the proposed scheme and two benchmark schemes. With increasing interference, the performance indicators worsen for all schemes; however, the proposed scheme provides the lowest average delivery times while spending comparable or less energy than the benchmarks.

\begin{table}[t]
    \centering
{
        \caption{
Normalized energy expenditure (EE) and delivery times (DT) averaged over node distances}
    \begin{tabular}{c|c|c|c|c|c|c}
       & \thead{Prop.}  & \thead{FSF-10} & \thead{FSF-11} & \thead{FSF-12} & \thead{GB-E} & \thead{GB-L} \\ \hline
    \small{EE (norm.)} & 11.6 &  13.4  & 16.3  & 26.5  &  8.7  & 10.7\\
    \small{DT (hours)} & 15.3  & 24.2 &   17.0 &  19.5 &   36.4 &  28.3
\\
    \vspace{-4mm} 
    \end{tabular}
        \label{tab:avgPerf}

}
\end{table}

\begin{table}[t]
    \centering
        \caption{
Performance under different traffic densities}
    \begin{tabular}{c|c|c|c|c|c|c}
      \multirow{2}{1em}{$\lambda_I$} & \multicolumn{3}{c|}{\small{Energy expenditure (norm.)}}  & \multicolumn{3}{c}{\small{Delivery time (hours)}} \\ 
      & \footnotesize{Prop.} & \footnotesize{FSF-11} & \footnotesize{GB-L} & \footnotesize{Prop.} & \footnotesize{FSF-11} & \footnotesize{GB-L} \\\hline
    $0.0005$ & 13.69 & 18.02 & 12.78 & 18.85  &  21.78 & 32.17\\
    0.001 & 14.50 & 18.45 & 12.9 & 20.04 & 22.57 & 34.21\\
    0.002 & 16.35 & 19.33 & 13.51 & 22.83 & 24.24 & 36.02\vspace{-3mm} 
    \end{tabular}
    \label{tab:traffic_impact}

\end{table}

\begin{figure}
\begin{subfigure}{0.5\textwidth}
  \centering
  % include first image
  \includegraphics[scale=0.55]{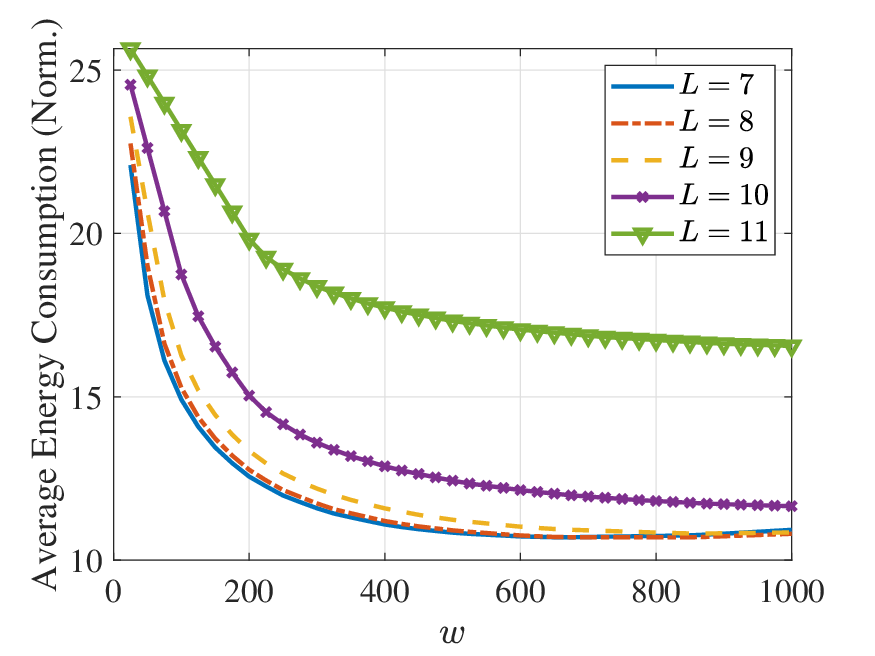}  
  \vspace{-3mm}
  \caption{Energy expenditure.}
  \label{fig:avg_energy_vs_w}
\end{subfigure}
\begin{subfigure}{0.5\textwidth}
  \centering
  % include second image
  \includegraphics[scale=0.55]{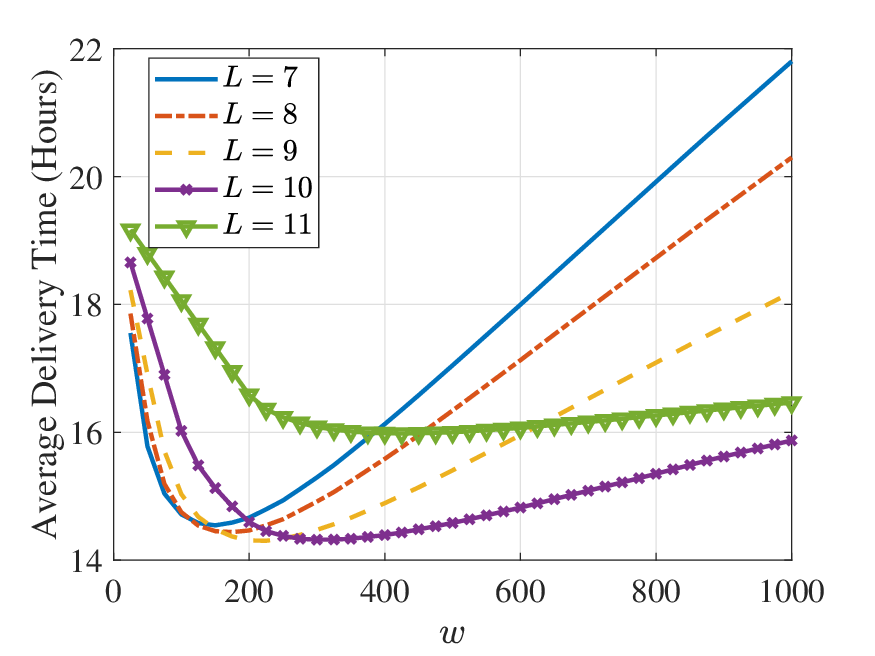}  
  \vspace{-3mm}
  \caption{Delivery time.}
  \label{fig:avg_latency_vs_w}
\end{subfigure}
\vspace{-4mm}
\caption{Performance as a function of $w$ (with $M\equals 12)$.}
\label{fig:vs_w}
\end{figure}

The impact of the number of transmissions per round, $w$, on the proposed scheme's performance is shown in Fig.~\ref{fig:vs_w}. We consider five SF sets that differ in their smallest SF $L$. For all sets, the largest SF is $M \equals 12$ so that the scheme can utilize the better receiver sensitivity of the higher SFs if required. We observe that for any $w$, making the SF set larger (i.e., reducing $L$) improves the average energy efficiency.  The inclusion of smaller SFs permits nearby recipients to receive fragments via shorter link-layer frames, thus saving energy at such recipients and reducing the energy consumption averaged over node distances. For the SF sets with $L \gteq 10$, the average energy consumption decreases monotonically with $w$ over the range shown. For $L \lteq 9$, although barely discernible, there is a slight convexity. In general, the larger the value of $w$, the more likely are nearby recipients to receive the entire update over low SFs (shorter frames, less energy expenditure) and distant recipients to make many failed reception attempts (high energy expenditure). These two competing phenomena lead to the  slight convexity in the average energy consumption. For the average delivery time, we observe a strong convex behavior. This is because as $w$ increases,  nearby recipients receive the update faster but  distant recipients must wait a long time for complete reception. Based on Fig.~\ref{fig:vs_w} and the application's requirement, a suitable $(w,L)$ pair may be chosen.   

We remark here that although our SF sets contain all SFs up to 12, the transmissions may not use all of them. For example, if all devices are very close to the gateway (e.g., \mbox{$\mathcal{R}_0 \lteq 100$ m}), plots similar to Fig.~\ref{fig:vs_w} would show that the optimal choice in terms of both energy consumption and delivery time is to use the SF set $\{7,8,\ldots,12\}$ with a very large $w$. This leads to the completion of the session with SF 7 transmissions alone.

\vspace{-2mm}

\section{Conclusion}
\label{sec:conlusion}
For efficient firmware updates over LoRaWAN, we devised a low-complexity  scheme that  employs multiple spreading factors sequentially to serve the needs of different end devices. The method can adjust the trade-off between the energy consumption and delivery-completion time by tuning a pair of design parameters. An analytical framework  is developed to determine the parameter values for the desirable operating point. Our contribution presents a solution to the shortcoming of existing schemes, which either provide energy efficiency at the cost of  high delivery delays, or vice versa, without an option to adjust this trade-off.

\appendices

% \vspace{-2mm}
\section{Derivation of $\overline{N}_s$}
\label{app:Ns}

Typically, $\overline{N}_s$ is slightly larger than $k$, although its exact value depends on the code. In general,
\begin{align}
    \overline{N}_s \equals  \sum_{m=k}^{\infty} m P_N(m),
\end{align}
where $P_N(m)$ is the probability that decoding requires the reception of exactly $m$ coded fragments. It is easy to see that
\begin{align}
    P_N(m) = \prod_{i=k}^{m-1} \hat{P}_f(i) \cdot (1-\hat{P}_f(m)),
\end{align}
where $\hat{P}_f(l)$ is the probability that the decoding attempt made upon receiving the $l$-th coded fragment fails, after having failed in each of the previous attempts. To find $\hat{P}_f(l)$, first define  $\E_l$ be the event that a collection of $l$ arbitrarily chosen coded fragments contains fewer than $k$ linearly independent combinations. Then $\hat{P}_f(l)$ can be written as
\begin{align} \label{eq:Pf_Gen}
\hat{P}_f(l) & \equals P(\E_l| \E_{l-1})  \equals  \frac{P(\E_{l}, \E_{l-1})}{P(\E_{l-1})} \equals  \frac{P(\E_{l})}{P(\E_{l-1})}.
\end{align}
For our illustrations, we assume the raptor code of~\cite{LGS07}. For this code, $P(\E_l) \approx 0.85 \times (0.567)^{l-k}$ for $l \geq k$~\cite{LGS07}. It follows that $\hat{P}_f(l) \equals 1$ for $l \lthan k$, $0.85$ for $l \equals k$, and $0.567$ for $l \gthan k$.

% \vspace{-3mm}
\section{Derivation of $\cF^{(\pr)}_{i}$ and  $\TSP_{i}^{(\fr)}$}
\label{app:S}
Consider a frame $\F$ employing SF $i$ sent by the gateway. Let $\F^{(\pr)}$ and $\F^{(\pl)}$ denote the preamble and the payload portions of $\F$, respectively.  Under the block-fading assumption, both  $\F^{(\pr)}$ and $\F^{(\pl)}$ are received with the same power \mbox{$R \equals \gamma_0 p_t A d_0^{-\alpha}$}, where $A$ is the fading coefficient. We first focus on the preamble $\F^{(\pr)}$. For its successful acquisition,  $R$ should be high enough to avoid both a detection failure and collision-induced failure. 
Conditioned on $A \equals a$,  the conditional acquisition probability is 
\begin{align}  \label{eq:S_FI_given_A}
    \mathcal{S}^{(\pr)}_{i}(d_0|A=a) &= \mathcal{D}^{(\pr)}(d_0|A=a) \mathcal{I}^{(\pr)}(d_0|A=a),
\end{align}
where $\mathcal{D}^{(\pr)}(d_0|A \equals a)$ is the probability of successful detection, that is, the probability that the power $R$ in $\F^{(\pr)}$ exceeds the receiver sensitivity, and $\mathcal{I}^{(\pr)}(d_0|A \equals a)$ is the conditional probability that $\F^{(\pr)}$ survives interference from other senders. It follows that $\mathcal{D}^{(\pr)}(d_0|A\equals a)$ 
is 1 if $\gamma_0 p_t a d_0^{-\alpha} \gteq \zeta_i$ and 0 otherwise. Using this fact and deconditioning on $A$, we obtain 
\begin{align}
\label{eq:S_FI_pr}
    \mathcal{S}_{i}^{(\pr)}(d_0) &= \int_{\zeta_i d_0^{\alpha} / \gamma_0 p_t}^{\infty} \mathcal{I}^{(\pr)}(d_0|A=a) \e^{-a} \dx{a}.    
\end{align}

To determine $\mathcal{I}^{(\pr)}(d_0|a)$, consider an interferer located at distance $D'$ from the intended recipient and using SF $j$ and transmitting $b'$-byte payloads. For this interferer to cause the loss of $\F^{(\pr)}$, it must send a frame $\F'$ such that (a) $\F^{(\pr)}$ and $\F'$ collide over the same channel and (b) the received power $R'$ in $\F'$ is such that $R/R' \lthan \xi_{s,j}$. 
For the collision criterion (a), recall that $\F^{(\pr)}$ has duration $l_i^{(\pr)}$, whereas $\F'$ has average duration $l^{(\fr)}_j(b')$. Using well known properties of the Poisson arrival process, the collision probability is given by $\lambda_f  (l_i^{(\pr)} \plus l^{(\fr)}_j(b')) / n_f$. Averaging over all possible $b'$, the average collision probability becomes $    C^{(\pr)}_{i,j} \equals \lambda_f  \left(l_i^{(\pr)}+\overline{l}_j\right)/n_f$.

For the power criterion (b), first condition on $D' \equals u$. Then the received power in $\F'$ is $R' \equals \gamma_0 p_t A' u^{-\alpha}$, where $A'$ is the fading coefficient. Since $\F^{(\pr)}$ has conditional received power $R \equals \gamma_0 p_t a d_0^{-\alpha}$, criterion (b) is satisfied with probability 
\begin{align}  \nonumber
    \mathcal{P}_{i,j}(d_0|D'=u) &= P(\gamma_0 p_t a d_0^{-\alpha}/\gamma_0 p_t A' u^{-\alpha} < \xi_{i,j}) \\  
    % &= P(A' >  \xi_{i,j}^{-1}a(d_0/u)^{-\alpha}) \\ 
    &= \e^{-\xi_{i,j}^{-1}a(d_0/u)^{-\alpha}}.
\end{align}

Then the conditional preamble loss probability due to one interferer can be written as 
\begin{align} \nonumber
\label{eq:Qpr}
    &Q^{(\pr)}_I(d_0|A=a) \\ \nonumber 
    &= \sum_{j=7}^{M} \eta_{j} C^{(\pr)}_{i,j} \int_{0}^{\R_I} \mathcal{P}_{i,j}(d_0|D'=u) f_D(u) \dx u\\ \nonumber
    &= \sum_{j=7}^{M} \eta_{j} C^{(\pr)}_{i,j} \int_{0}^{\R_I} \e^{-\xi_{i,j}^{-1}a(d_0/u)^{-\alpha}} f_D(u) \dx u\\ \nonumber
    &= \frac{2}{\R_I^2}  \sum_{j=7}^{M} \eta_{j}C^{(\pr)}_{i,j}\int_{0}^{\R_I} u\e^{-\xi_{i,j}^{-1}a(d_0/u)^{-\alpha}} \dx u\\ 
    % &= \frac{2}{\R_I^2} \int_0^{\R_I} u\e^{-\beta u^{\alpha}} \dx u\\ 
    &=\frac{2}{\alpha\R_I^2}\sum_{j=7}^{M}\frac{\eta_{j} C^{(\pr)}_{i,j}}{\beta_{j}^{2/\alpha} }\gamma(2/\alpha,\beta_{j} \R_I^\alpha),
\end{align}
where \mbox{$\beta_j \equals \xi_{i,j}^{-1}ad_0^{-\alpha}$} and $\gamma(i,x) \equals \int_{0}^{x}t^{i-1}\mathrm{e}^{-t}\dx{t}$.
Since there are $n$ interferers, the conditional probability that none of them causes the loss of $\F^{(\pr)}$ is 
\begin{align} \label{eq:I_pr}
\mathcal{I}^{(\pr)}(d_0|A=a) = (1 - Q^{(\pr)}_I(d_0|A=a))^{n}.    
\end{align}
Substitution of~\eqref{eq:I_pr} in \eqref{eq:S_FI_pr} yields $\TSP_{i}^{(\pr)}$, and thus, $\cF_{i}^{(\pr)}$.

The derivation of $ \TSP_{i}^{(\fr)}$ is identical, except that the  collision probability is given by
$C^{(\fr)}_{i,j} = \lambda_f   \left(l_i(b_k)+\overline{l}_j\right)/n_f$.  

For $\cF_{i}^{(\pl)}$, note that $\mathcal{S}_{i}^{(\fr)}$ is  the joint probability of preamble acquisition and payload reception. From Bayes rule,
    \mbox{$\mathcal{S}_{i}^{(\pl)} = {\mathcal{S}_{i}^{(\fr)}}/{\mathcal{S}_{i}^{(\pr)}}$}; hence, \mbox{$\cF_{i}^{(\pl)} = 1 - {\mathcal{S}_{i}^{(\fr)}}/{(1-\cF_{i}^{(\pr)})}$}

\balance
\bibliographystyle{IEEEtran}
\bibliography{refs.bib}

% \begin{IEEEbiography}[{\includegraphics[width=1in,height=1.25in,clip,keepaspectratio]{Author Photos/Siddhartha_Borkotoky.jpg }}]{Siddhartha S. Borkotoky}
% received the B.Tech. degree in electronics and instrumentation engineering from the National Institute of Technology Rourkela in 2008 and the Ph.D. degree in electrical engineering from Clemson University in 2017. From 2008 to 2010, he was a research assistant with the Indian Institute of Technology Guwahati. He was a post-doctoral fellow with Clemson University from 2017 to 2018, and a Senior Researcher with Lakeside Labs GmbH, Klagenfurt, Austria from 2018 to 2019. He is currently an assistant professor with the Indian Institute of Technology Bhubaneswar. His research interests include  wireless sensor networks, cooperative communications, and application-layer coding techniques. 
% \end{IEEEbiography}

\end{document}